\documentclass[letterpaper, 10 pt, conference]{ieeeconf}  

\IEEEoverridecommandlockouts                              

\usepackage{graphicx}

\usepackage{stfloats}
\usepackage{float}
\usepackage{subfigure}
\usepackage{svg}
\usepackage{enumerate}
\usepackage{algorithm}
\usepackage{algpseudocode}

\usepackage{amsmath, amsfonts, amssymb, amsthm}

\newtheorem{theorem}{Theorem}
\newtheorem{lemma}{Lemma}

\newtheorem{problem}{Problem}
\newtheorem{remark}{Remark}

\title{\LARGE \bf
Harnessing Data for Accelerating Model Predictive Control by Constraint Removal}

\author{Zhinan Hou, Feiran Zhao, Keyou You
\thanks{*The research was supported by National Science and Technology Major Project of China (2022ZD0116700) and National Natural Science Foundation of China (62033006, 62325305). }
\thanks{Z. Hou, F. Zhao, and K. You are with the Department of Automation and BNRist, Tsinghua University, Beijing 100084, China. e-mail: hzn22@mails.tsinghua.edu.cn, zhaofr18@mails.tsinghua.edu.cn, youky@tsinghua.edu.cn.}
}

\begin{document}

\maketitle
\thispagestyle{empty}
\pagestyle{empty}

\begin{abstract}
	 Model predictive control (MPC) solves a receding-horizon optimization problem in real-time, which can be computationally demanding when there are thousands of constraints. To accelerate online computation of MPC, we utilize data to adaptively remove the constraints while maintaining the MPC policy unchanged. Specifically, we design the removal rule based on the Lipschitz continuity of the MPC policy. This removal rule can use the information of historical data according to the Lipschitz constant and the distance between the current state and historical states. In particular, we provide the explicit expression for calculating the Lipschitz constant by the model parameters. Finally, simulations are performed to validate the effectiveness of the proposed method.
\end{abstract}

\section{Introduction}
Model predictive control (MPC) is a popular optimal control technique and has many successful applications, including autonomous driving \cite{cesari2017scenario}, industrial process \cite{qin2003survey}, and robotic manipulation \cite{best2016new}. By solving online a receding-horizon optimization problem, the MPC computes the control input in real-time that ensures the satisfaction of system constraints. However, solving an optimization problem per timestep can be computationally demanding even for linear dynamical systems, especially when there are thousands of constraints \cite{van2000model}. Thus, reducing the computation burden has become an important problem for both theoretical and practical researchers.

Manifold approaches to accelerating the online computation include model reduction \cite{hovland2006mpc}, explicit MPC \cite{bemporad2002explicit}, tailored solvers \cite{jerez2011condensed}, and constraint removal \cite{ardakani2014acceleration}. Recently, there has been an increasing interest in the constraint removal approach as it can utilize online data to further reduce the computation time \cite{nouwens2021constraint}. The constraint removal method first pre-determines  unnecessary (inactive) constraints of the receding-horizon optimization problem by exploiting local properties of the optimization problem, such as reachable sets \cite{nouwens2021constraint}, region of activity \cite{jost2013accelerating}, and contraction properties of the cost function \cite{jost2015online}. Then, it removes these constraints to reduce the scale of the optimization problem, while maintaining the solution unchanged. Very recently, Nouwens et al. \cite{nouwens2023constraint} proposes a general system-theoretic principle to design the constraint removal rule using the online closed-loop state-input data, leading to the constraint-adaptive MPC (ca-MPC) framework. While the ca-MPC enables the use of system-theoretic properties (e.g., reachability and optimality), it is conservative as only the data at current time is utilized. In fact, the historical data should also be taken into account in some applications. For example, the racing car routinely runs on the same track, and using data from past iterations may further improve the performance of the constraint-removal method.

In this paper, we propose a new constraint removal method for the ca-MPC of linear systems, which uses the historical state-input data to further accelerate the online computation. Instead of using reachability and optimality of the receding-horizon problem \cite{nouwens2023constraint}, we exploit the Lipschitz continuity of the MPC policy in the state to design the removal rule. However, an explicit Lipschitz constant is challenging to obtain and is absent in the literature. Fortunately, by estimating the KKT solution of the receding-horizon problem, we are able to provide an explicit expression for the Lipschitz constant. In particular, the Lipschitz constant can be computed offline using only the model parameters. By leveraging the Lipschitz continuity, the proposed constraint removal rule can utilize the historical state data that is ``close" to the current system state.  Finally, simulations are performed to validate the effectiveness of the proposed method.

Our result on the Lipschitz constants of the MPC policy has independent interests of its own. The Lipschitz constant is beneficial in various realization of MPC including robustness against disturbance \cite{scokaert1997discrete} and neural network-based approximation \cite{9928332}. However, an explicit Lipschitz constant is challenging to obtain. While recent works estimate the Lipschitz constant via enumerating potential active sets \cite{darup2017maximal} or a mixed-integer linear program \cite{9928332}\cite{teichrib2023efficient}, they are time-consuming and lack an explicit expression. In contrast, we provide the first explicit Lipschitz constant that can be efficiently computed using model parameters. 

The rest of this paper is organized as follows. The notation is given in the remainder of this section. Section \ref{problem} introduces the linear MPC and formulates the problem. We provide an explicit expression of the Lipschitz constant and present the ca-MPC scheme in Section \ref{method}. We illustrate the effectiveness of the proposed scheme with simulation in Section \ref{exp} and give a conclusion in Section \ref{con}. 

\textit{Notation.} We denote the set of real number by $\mathbb{R}$, the set of $n$-dimensional real-valued vectors by $\mathbb{R}^n$ and the set of $n \times m$-dimensional real-valued matrices by $\mathbb{R}^{n \times m}$. Given a matrix $A \in \mathbb{R}^{m \times n}$, $A^T$ denotes its transpose, $A_j$ denotes its $j$-th row and so is the vector. $\lambda_i(A)$ is the i-th largest eigenvalue of $A$. Specially, we define $\lambda_{min}(A)$ as the smallest eigenvalue of $A$. For two vectors $a,b$, we denote element-wise multiplication by $a \circ b$ . For any set of indices $\mathbb{I} \subseteq \{1,...,m\}$, $A_{\mathbb{I}}$ denotes the submatrix obtained by selecting the rows indicated in $\mathbb{I}$. $\text{card}(S)$ is denoted as the number of elements in the set $S$. $\mathbb{N}_{[a,b]} = \{ n \in \mathbb{N} \ | \ a \le n \le b\}$. For two sets $A, B$, we define $A - B = \{ x \ | \ x \in A , x\notin B\}$. 

\section{Problem formulation and background} \label{problem}

In this section, we introduce the linear system to be controlled and the linear MPC setup (see Section \ref{problem_A}). Furthermore, we introduce the constraint-adaptive MPC and formulate our problem in Section \ref{problem_B}.

\subsection{Linear MPC} \label{problem_A}
We consider a discrete-time linear time-invariant (LTI) system:
\begin{equation}
	x_{k+1} = Ax_k + Bu_k, \label{2_1}
\end{equation}
where $x_k \in \mathbb{R}^n$ and $u_k \in \mathbb{R}^m$ denote the states and the inputs, respectively, at discrete time $k \in \mathcal{N}$. $A \in \mathbb{R}^{n \times n}$ and $B \in \mathbb{R}^{m \times m}$ are known system matrices. The system is subject to polyhedral state and input constraints:
\begin{equation}
	(x_k, u_k) \in \mathbb{L} = \{(x,u) \ | \ Cx+Du \le E\}, \label{2_2}
\end{equation}
where $C \in \mathbb{R}^{c \times n}, \ D \in \mathbb{R}^{c \times m}, \ E \in \mathbb{R}^{c}$ and origin is in the interior of $\mathbb{L}$.

Based on the system dynamic \eqref{2_1} and the constraints \eqref{2_2}, an MPC over a finite time horizon of length $T \le 1$, given the state $x_k$ at time $k \in \mathbb{N}$, can be formulated as
\begin{equation}
	\begin{aligned}
		\min_{u_{.|k}} \ & J(x_{.|k}, u_{.|k}), \\
		\text{s.t.}    \ & x_{t+1|k} = Ax_{t|k} + Bu_{t|k}, \\
		& Cx_{t|k} + Du_{t|k} \le E, \\
		& C_Tx_{N|k} \le E_T, \\
		& x_{0|k} = x_k, \\
		& t = 0,1,...,N-1, 	\label{2_3}
	\end{aligned}
\end{equation}
where 
\begin{equation}
	J(x_{.|k}, u_{.|k}) := \sum^{N-1}_{t=0} \frac{1}{2}(x_{t|k}^TQx_{t|k} + u_{t|k}^TRu_{t|k}) + \frac{1}{2}x_{N|k}^TPx_{N|k}. \label{2_4}
\end{equation}
Here, $x_{.|k}, u_{.|k}$ denote the predicted state and the predicted input. $J(x_{.|k}, u_{.|k})$ is composed of the stage cost and the terminal cost where the weights $Q, R, P$ are symmetric positive definite matrices. The constraint $\mathbb{X}_T = \{x \ | \ C_Tx \le E_T\}$ is the terminal constraint.

Let $u^*_{.|k}$ be the solution of \eqref{2_3}. Then, the first input is applied to the system:
\begin{equation}
	u_{MPC}(k) = u^*_{0|k}. \label{2_5}
\end{equation}

The optimization \eqref{2_3} can be expressed as a standard multiparameter quadratic programming (mp-QP) form by regarding $x$ as the parameter vector:
\begin{subequations}
	\begin{align}
		\min_z \ & \frac{1}{2}z^THz + x^TFz, \label{2_7_1}\\ 
		\text{s.t.} \ & z \in \mathbb{Z}(x) := \{z \ | \ Gz \le W + Sx \}, \label{2_7_2}
	\end{align} \label{2_7}
\end{subequations}
where $x \in \mathbb{R}^n$ is the current state, $H \in \mathbb{R}^{n_z \times n_z}, F \in \mathbb{R}^{n \times n_z}, G \in \mathbb{R}^{n_c \times n_z}, W \in \mathbb{R}^{n_c}, S \in \mathbb{R}^{n_c \times n}$ are parameter matrices which are obtained from $Q, R, P, A, B, C, D, E, C_T, E_T$. We denote an optimal solution to \eqref{2_7} by $z^*(x)$. The set of indices of the active constraints responding to $x$ is denoted by $\mathcal{A}(x)$.

If mp-QP has many constraints in \eqref{2_7_2}, then it is challenging and time-consuming to solve this optimization problem. One approach to reduce computational burden is to remove some redundant constraints.

\subsection{Constraint-adaptive MPC} \label{problem_B}

\begin{figure}[t]
	\centering
	\includegraphics[width=6cm]{{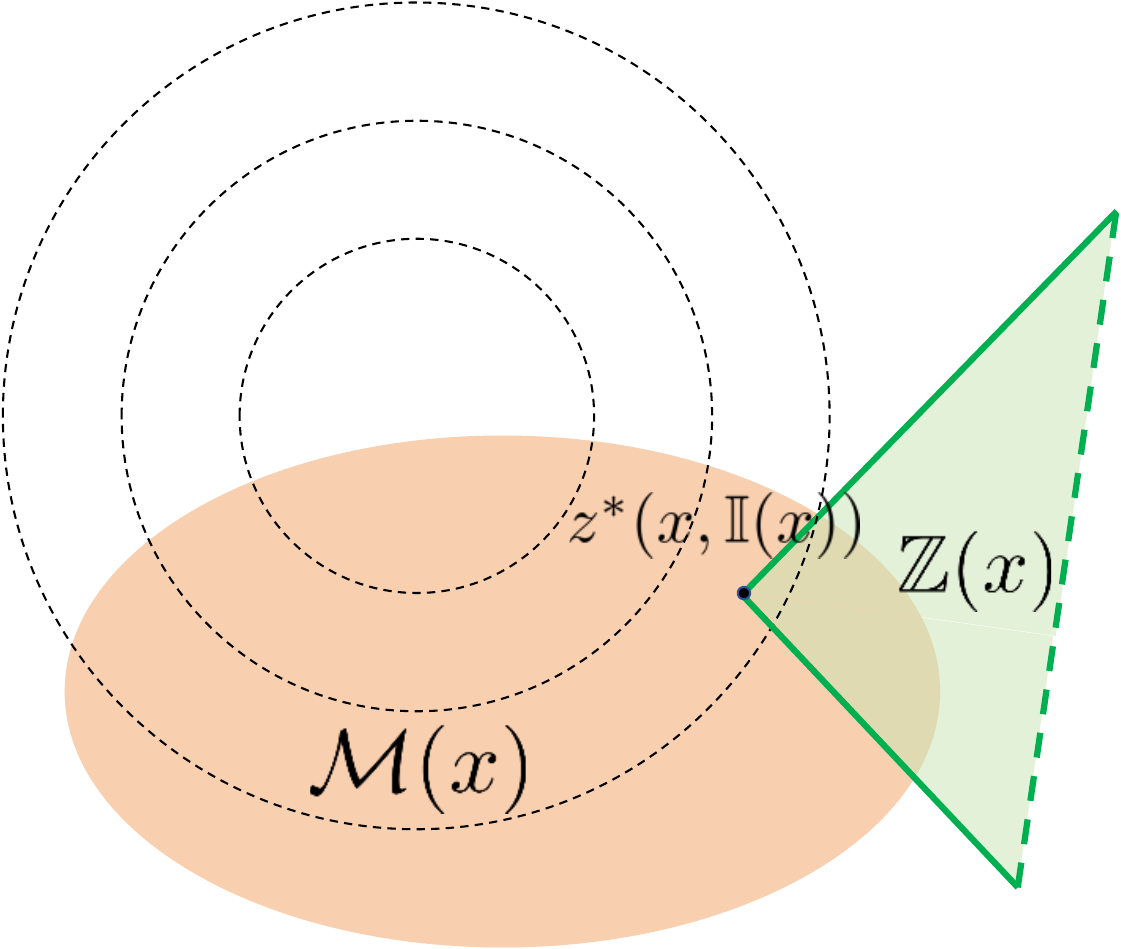}}
	\caption{The illustrations of Lemma \ref{outer}. The black dashed lines denote level sets of the cost function. The orange region denotes $\mathcal{M}(x)$. The green solid and dashed lines denote the constraints specified by $\mathbb{I}(x)$ and $\mathbb{C}(x)$, respectively. The left half-space of the green dashed line is the set $\mathbb{Z}(x,\mathbb{C}(x))$.}
	\label{fig:approximation}
\end{figure}

Nouwens et al. \cite{nouwens2023constraint} proposed a online constraint removal framework for linear system, called constraint-adaptive MPC. Specially, his method introduced the reduced MPC problem which have fewer constraints compared to the original MPC problem. The mp-QP version of the reduced MPC problem is formulated as follows:
\begin{subequations}\label{2_8}
	\begin{align}
		\min_z \ & \frac{1}{2}z^THz + x^TFz, \label{2_8_1}\\ 
		\text{s.t.} \ & z \in \mathbb{Z}(x, \mathbb{I}(x)) = \{z \ | \ G_jz \le W_j + S_jx, j\in \mathbb{I}(x)\} ,  \label{2_8_2}
	\end{align} 
\end{subequations}
where we define the mapping $\mathbb{I}: \mathbb{R}^n \rightrightarrows \mathbb{N}_{[1,n_c]}$ from the state space to the indices set of reduced constraints. $z^*(x, \mathbb{I}(x))$ denote the optimal solution of \eqref{2_8}.

To guarantee that the optimal solution to \eqref{2_8} is the same as that to \eqref{2_7},  Nouwens et al. \cite{nouwens2023constraint} build upon the theorem as follows:
\begin{lemma}[Theorem 1, \cite{nouwens2023constraint}] \label{outer}
	Consider the original and reduced MPC optimization problem \eqref{2_7} and \eqref{2_8}. Let the set-valued mapping $\mathbb{C}: \mathbb{R}^n \rightrightarrows \mathbb{N}_{[1,n_c]}$ denote the set of indices of removed constraints, i.e., $\mathbb{C}(x) = \mathbb{N}_{[1,n_c]} - \mathbb{I}(x)$. If there exists a mapping $\mathcal{M}: \mathbb{R}^n \rightrightarrows \mathbb{R}^{n_z}$ such that for all $x \in \mathbb{R}^n$,
	\begin{subequations}
		\begin{align}
			& z^*(x, \mathbb{I}(x)) \in \mathcal{M}(x), \label{2_9_1} \\
			& \mathcal{M}(x) \subset \mathbb{Z}(x, \mathbb{C}(x)). \label{2_9_2}
		\end{align}
	\end{subequations}
	Then, it holds that $z^*(x, \mathbb{I}(x)) = z^*(x)$.
\end{lemma}

In Lemma \ref{outer}, $\mathcal{M}(x)$ is regarded as the outer approximation of $z^*(x,\mathbb{I}(x))$. This approximation can be obtained by some information about the optimal solution (e.g., reachability and optimality \cite{nouwens2023constraint}). 

The way to apply the Lemma \ref{outer} is to first construct the outer approximation $\mathcal{M}(x)$ of $z^*(x, \mathbb{I}(x))$ for all possible choices of $\mathbb{I}$. Then, the indices in $\mathbb{C}(x)$ are selected such that \eqref{2_9_2} is satisfied. Finally, constraints in $\mathbb{C}(x)$ are removed to formulate the reduced mp-QP problem \eqref{2_8}. As shown in Fig. \ref{fig:approximation}, $\mathcal{M}(x)$ is in the half-space formed by $\mathbb{C}(x)$ and contained by $\mathbb{Z}(x, \mathbb{C}(x))$. In this case, the constraints specified by $\mathbb{C}(x)$ can be removed directly without changing the optimal solution, i.e. $z^*(x, \mathbb{I}(x)) = z^*(x)$. Thus, the design of outer approximation $\mathcal{M}(x)$ is essential, and a good approximation leads to efficient constraint removal. In this paper, we are interested in the following problem:

\begin{problem}
	Design an approximation $\mathcal{M}(x)$ in Lemma \ref{outer} and improve the constraints removal.
\end{problem}

\section{Exploiting historical data for ca-MPC} \label{method}

In this section, we use the historical data to design the outer approximation $\mathcal{M}(x)$. This is motivated by an observation that the optimal solution for mp-QP is Lipschitz continuous. If the current state $x$ is close to the previous state $\tilde{x}$, then the optimal solution $z^*(x)$ of the current state is in a circle centered around that $z^*(\tilde{x})$ of which radius is $\kappa_{max} \Vert x - \tilde{x}\Vert_2$. This information can be used to design the outer approximation $\mathcal{M}(x)$, then evaluate and remove redundant constraints. Specifically, we provide an explicit form of the Lipschitz constant of linear MPC in Section \ref{S_3_1}. Based on this, we design a constraint removal mechanism that utilizes historical data and present our MPC scheme in Section \ref{S_3_2}.

\subsection{The Lipschitz constant of reduced mp-QP} \label{S_3_1}
Consider a mp-QP problem with constraints selected by $\mathbb{I}$ which has no relation with the state x compared to \eqref{2_8}: 
\begin{subequations}
	\begin{align}
		\min_z \ & \frac{1}{2}z^THz + x^TFz, \label{3_2_1}\\ 
		\text{s.t.} \ & z \in \mathbb{Z}_\mathbb{I}(x) = \{z \ | \ G_\mathbb{I}z \le W_\mathbb{I} + S_\mathbb{I} x \}. \label{3_2_2}
	\end{align} \label{3_2}
\end{subequations}
Here, we denote the optimal solution of \eqref{3_2} by $z^*_{\mathbb{I}}(x)$. For each $\mathbb{I}$, there exists a $\kappa_{\mathbb{I}}$  \cite{bemporad2002explicit} such that
\begin{equation}
	\Vert z_\mathbb{I} ^*(x_1) - z_\mathbb{I} ^*(x_2) \Vert \le \kappa_\mathbb{I} \Vert x_1 - x_2 \Vert. \label{3_3}
\end{equation}

The Lipschitz constant $\kappa_\mathbb{I}$ depends on $\mathbb{I}$. It is hard to compute the Lipschitz constant $\kappa_\mathbb{I}$ unless the explicit law is known \cite{darup2017maximal}. In fact, the explicit law is intractable for large-scale system. In our cases, we need to calculate the maximum Lipschitz constant of mp-QP in different $\mathbb{I}$, i.e., $\kappa_\text{max} = \max_{\mathbb{I} \subseteq \mathbb{N}_{[1,n_c]}} \kappa_\mathbb{I}$. As far as we know, no methods can calculate this constant efficiently. To address this problem, we provide an explicit form for $\kappa_\text{max}$ with very low computational cost as follows:

\begin{theorem} \label{continuity}
	For any $\mathbb{I} \subseteq \mathbb{N}_{[1,n_c]}, x_1,x_2 \in \mathbb{R}^n$, it holds that
	\begin{equation}
		\Vert  z_\mathbb{I}^*(x_1) - z_\mathbb{I}^*(x_2) \Vert_2 \le \kappa_\text{max} \Vert x_1 - x_2 \Vert_2, \label{3_4}
	\end{equation}
	with 
	\begin{equation}
		\begin{aligned}
			\kappa_\text{max} = 
			&\Vert H^{-1} F^T\Vert_2 + \frac{1}{\min_j G_jH^{-1}G_j^T}\Vert H^{-1}G^T\Vert_2 \\
			&\times \Vert S + GH^{-1}F^T\Vert_2. \label{3_5}
		\end{aligned}	
	\end{equation}
\end{theorem}

\begin{remark}
	This Lipschitz constant is finite because $H^{-1}$ is positive definite matrix and $G_jH^{-1}G_j^T > 0$. Also, this constant can be offline computed from the model parameters.
\end{remark}

Theorem \ref{continuity} provides an efficient computing method for the Lipschitz constant. Its computation includes matrix multiplication and matrix norm. Let us analyze the computation complexity of the matrix norm. Here, we are interested in the mp-QP that exists many constraints, i.e., $n_c \gg n + n_z$. For a matrix $X \in \mathbb{R}^{a \times b}$, its matrix norm $\Vert X \Vert_2 = \sqrt{\lambda_1(XX^T)} = \sqrt{\lambda_1(XX^T)}$. The computation complexity per step of $\lambda_1(XX^T)$ and $\lambda_1(X^TX)$ with power iteration methods are $\mathcal{O}(a^2)$ and $\mathcal{O}(b^2)$, respectively. So the computation complexity of $\Vert X \Vert_2$ is $\mathcal{O}(a^2)(a < b)$ or $\mathcal{O}(b^2)(a > b)$. The dimensions of three matrix in \eqref{3_5} to compute norm are $n_z \times n$, $n_z \times n_z$ and $n_c \times n$. Although $n_c \gg n + n_z$, the computation complexity of matrix norm in \eqref{3_5} is $\mathcal{O}(n^2 + n_z^2)$ and are not influenced by the number $n_c$ of constraints.

This upper bound in Theorem \ref{continuity} may be not close. To decrease the gap, we can utilize some techiques to find a smaller value $\kappa_{\text{max}}$. Let $\Phi \in R^{n_c \times n_c}$ be a full-rank matrix. We substitute $G, W, S$ with $\Phi G, \Phi W, \Phi S$ in \eqref{2_7_2} and the feasible region and the optimal solution are not changed. The variant of \eqref{3_5} is obtained:
\begin{equation}
	\begin{aligned}
		\hat{\kappa}_\text{max} 
		&= \Vert H^{-1} F^T\Vert_2 + \frac{1}{\min_j (\Phi G)_jH^{-1}(\Phi G)_j^T}\Vert H^{-1}G^T\Phi^T \Vert_2 \\
		& \times \Vert \Phi S + \Phi GH^{-1}F^T\Vert_2. \label{3_6}
	\end{aligned}	
\end{equation}

Equation \eqref{3_6} can be used as the Lipschitz constant in \eqref{3_4}. We can select the appropriate matrix $\Phi$ to find a smaller value $\hat{\kappa}_\text{max}$ than $\kappa_\text{max}$ in \eqref{3_5}. In Section \ref{exp}, we provide a empirical approach to choose $\Phi$. 

\subsection{The outer approximation $\mathcal{M}(x)$} \label{S_3_2}
We now utilize the Theorem \ref{continuity} to construct the outer approximation $\mathcal{M}(x)$. Then, we propose the ca-MPC scheme based on the mapping $\mathbb{I}$ computed by $\mathcal{M}(x)$.

We present the key lemma used in the design of $\mathcal{M}(x)$:
\begin{lemma} \label{data}
	For any $x_1, x_2 \in \mathbb{R}^n$, if $\mathcal{A}(x_2) \subseteq \mathbb{I}(x_1)$, we have
	\begin{equation}
		\Vert z^*(x_1, \mathbb{I}(x_1)) - z^*(x_2) \Vert_2 \le \kappa_\text{max} \Vert x_1 - x_2\Vert_2. \label{3_7}
	\end{equation}
\end{lemma}

\begin{proof}
	Let $\mathbb{I} = \mathbb{I}(x_1)$ in Theorem \ref{continuity}:
	\begin{equation}
		\Vert z^*(x_1, \mathbb{I}(x_1)) - z^*(x_2, \mathbb{I}(x_1)) \Vert_2 \le \kappa_\text{max} \Vert x_1 - x_2 \Vert_2. \label{3_8}
	\end{equation}

    We denote the objective function as $V(z) = \frac{1}{2}z^THz + x^T_2Fz$.
    Because $\mathcal{A}(x_2) \subseteq \mathbb{I}(x_1) \subseteq \mathbb{N}_{[1,n_c]}$, we have
    \begin{equation}
    	V(z^*(x_2, \mathcal{A}(x_2))) \le V(z^*(x_2, \mathbb{I}(x_1))) \le V(z^*(x_2)). \nonumber
    \end{equation}
    Then, it follows from $V(z^*(x_2, \mathcal{A}(x_2))) = V(z^*(x_2))$ that
    \begin{equation}
    	V(z^*(x_2, \mathcal{A}(x_2))) =  V(z^*(x_2, \mathbb{I}(x_1))) = V(z^*(x_2)). \label{3_10}
    \end{equation}
    $z^*(x_2, \mathbb{I}(x_1)),z^*(x_2) \in \mathbb{Z}(x_2, \mathcal{A}(x_2))$ and are optimal in \eqref{2_8} where let $x= x_2$ and $\mathbb{I}(x) = \mathcal{A}(x_2)$ according to \eqref{3_10}. Since this mp-QP have a unique optimal solution, so we have
    \begin{equation}
    	z^*(x_2, \mathbb{I}(x_1)) = z^*(x_2). \label{3_11}
    \end{equation}
    
    Substituting \eqref{3_11} into \eqref{3_8} and the proof is finished.
\end{proof}

In Lemma \ref{data}, we consider $x_1 = x$ as the current state and  $x_2=\tilde{x}$ as a historical state. Let reduced constraint indices $\mathbb{I}(x)$ satisfy $\mathcal{A}(\tilde{x}) \subseteq \mathbb{I}(x)$:
\begin{equation}
	z^*(x, \mathbb{I}(x)) \in \mathcal{M}(x), \label{3_12}
\end{equation}
where
\begin{equation}
	\mathcal{M}(x) = \{z \ | \ \Vert z - z^*(\tilde{x}) \Vert_2 \le \kappa_\text{max} \Vert x - \tilde{x} \Vert_2 \}. \label{3_13}
\end{equation} 

Equation \eqref{3_13} indicates that $\mathcal{M}(x)$ is a sphere set. The term $\Vert x - \tilde{x} \Vert_2$ in \eqref{3_13} is the distance between the current state and historical state. This distance is expected to be small so that $\mathcal{M}(x)$ can be contained in many constraints of $\mathbb{Z}(x)$. Based on this distance, we can evaluate if the historical state can provide good information for $\mathbb{C}(x)$ and can help remove some redundant constraints.  As a result, the historical data which have the smallest distance  $\Vert x - \tilde{x} \Vert_2$ is used to design $\mathcal{M}(x)$ and construct $\mathbb{C}(x)$.

To select the removal constraints $\mathbb{C}(x)$, we utilize the sphere half-space intersection check to evaluate if the sphere set $\mathcal{M}(x)$ is entirely constained in the half-space specified by the constraint in $\mathbb{Z}(x)$. This check leads to our removal rule:

\begin{equation}
	\begin{split}
		\mathbb{C}(x) = \{j \in \mathbb{N}_{[1,n_c]} \ | \ & \kappa_{\text{max}} \Vert x - \tilde{x} \Vert_2 \Vert G_j \Vert_2 < W_j + S_j x \\ 
		& - G_j z^*(\tilde{x}) \} - \mathcal{A}(\tilde{x}). \label{3_14}
	\end{split}
\end{equation}

In Algorithm \ref{alg:mpc}, we present the removal process and present our overview of the proposed MPC scheme. Theorem \ref{results} shows that our removal rule \eqref{3_14} maintains the trajectory unchanged.

\begin{algorithm}[t]
	\caption{implementation of ca-MPC using historical data} \label{alg:mpc}
	\begin{algorithmic}[1]
		\State $k = 0$; \label{a_1}
		\State Initialize historical data $D$; \label{a_2}
		\State $\kappa_{\text{max}} \gets \eqref{3_5} \text{ or } \eqref{3_6}$; \label{a_3}
		\While{True}
		\State Measure $x_k$; \label{a_4}
		\State{Find $\tilde{x}$ with the smallest distance $\Vert x_k - \tilde{x}\Vert_2$ from historical data $D$} \label{a_5_1}
		\State $\mathbb{C}(x_k) \gets \eqref{3_14}$ with data $(\tilde{x}, z^*(\tilde{x}), \mathcal{A}(\tilde{x}))$; \label{a_5_2}
		\State $\mathbb{I}(x_k) = \mathbb{N}_{[1, n_c]} - \mathbb{C}(x_k)$; \label{a_6}
		\State $z^*(x_k), \mathcal{A}(x_k) \gets \eqref{2_8}$ with $x = x_k $ and $ \mathbb{I}(x_k)$; \label{a_7}
		\State Store data $(x_k, z^*(x_k), \mathcal{A}(x_k))$ in $D$; \label{a_8}
		\State Apply control input from $z^*$ into plant; \label{a_9}
		\State $k \gets k+1$; \label{a_10}
		\EndWhile
	\end{algorithmic}
\end{algorithm}

\begin{theorem} \label{results}
    Given a original and reduced mp-QP problem \eqref{2_7} and \eqref{2_8}, the Lipschitz constant $\kappa_{\text{max}}$ \eqref{3_5}, the removal rule $\mathbb{C}(x)$ \eqref{3_14}, the state trajectory under Algorithm \ref{alg:mpc} is the same with closed-loop trajectory under the policy \eqref{2_7}. 
\end{theorem}

\begin{proof}
	We prove this Theorem by induction.
	
	At the k step of the Algorithm \ref{alg:mpc}, it follows from line \ref{a_5_2}-\ref{a_6} that
	\begin{equation} 
		\mathcal{A}(\tilde{x}) \subset \mathbb{I}(x_k). \label{3_15}
	\end{equation}
	
	Then it follows from Lemma \ref{data} that
	\begin{equation}
	    z^*(x_k, \mathbb{I}(x_k)) \in \mathcal{M}(x_k). \label{3_16}
	\end{equation} 
	with 
	\begin{equation}
		\mathcal{M}(x_k) = \{ z \ | \ \Vert z - z^*(\tilde{x})) \Vert_2 \le \kappa_\text{max} \Vert x_k - \tilde{x}\Vert_2 \}. \label{3_17}
	\end{equation}

	For each $j \in \mathbb{C}(x_k)$, it follows from line \ref{a_5_2} that
	\begin{equation}
		\kappa_{\text{max}} \Vert x - \tilde{x}\Vert_2 < \frac{W_j + S_j \tilde{x} - G_j z^*(\tilde{x})}{\Vert G_j \Vert_2}. \label{3_18}
	\end{equation}
	
	We know that the distance between $z^*(\tilde{x})$ and plane $Z_j = \{v \ | \ G_jv = W_j + S_j \tilde{x} \}$ is $\frac{W_j + S_j \tilde{x} - G_j z^*(\tilde{x})}{\Vert G_j \Vert_2}$. It follow from \eqref{3_16}\eqref{3_17}\eqref{3_18} that $\mathcal{M}(x_k)$ does not intersect with the plane $Z_j$. Also we know $z^*(\tilde{x}) \in Z_j$. Then $\mathcal{M}(x_k) \subset Z_j$. As a result, it holds that
	\begin{equation}
		\mathcal{M}(x_k) \subset \cap_{j \in \mathbb{C}(x_k)} Z_j = \mathbb{Z}(x, \mathbb{C}(x_k)). \label{3_19}
	\end{equation} 
	
	According to Lemma \ref{outer}, combining \eqref{3_16}, \eqref{3_19} with \eqref{3_15} yields
	\begin{equation}
		z^*(x_k, \mathbb{I}(x_k)) = z^*(x_k).  \nonumber 
	\end{equation}
	which can complete the proof by induction.
	
\end{proof}

Theorem \ref{results} indicates that our algorithm exploits the historical data to accelerate the optimization without changing the state trajectory. Compared to solving the optimization problem \eqref{2_7} with a large number of constraints, 
the computation in line \ref{a_5_1} of Algorithm \ref{alg:mpc} is low unless the historical data is very large. Even though this computation is limited by the number of historical data, some techiques can be used to overcome this problem. For example, the redundant data can removed when distance between two historical data is small. 

In Algorithm \ref{alg:mpc}, when data set $D$ is empty, we can initialize with one data $(x, z^*(x), \mathcal{A}(x)) = (0, 0 , \emptyset)$. In \eqref{3_14}, The terms such as $\Vert G_j \Vert_2$ can be calculated offline which can accelerate the construction of removal constraints.

\section{Simulations} \label{exp}
In this section, we apply the proposed method to the double integrator system. The simulations are conducted in MATLAB.

\begin{figure}[t]
	\centering
	\includegraphics[width=8cm]{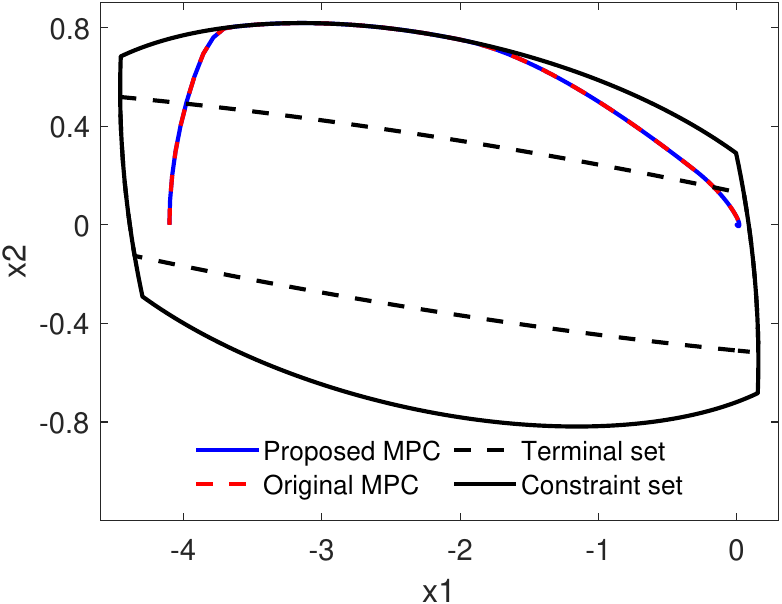}
	\caption{The closed-loop state trajectory using the original MPC policy and the proposed ca-MPC policy.}
	\label{fig:trajectory}
\end{figure}

Consider the double integrator system:
\begin{equation}
	x_{k+1} = \begin{bmatrix}
		1 & 0.1 \\ 0 & 1
	\end{bmatrix} x_k + \begin{bmatrix}
		0.005 \\ 0.1
	\end{bmatrix} u_k,  \nonumber
\end{equation}
where the input is constrained by $\vert u \vert \le 1$. The state constraints are
\begin{subequations}
	\begin{align}
		& (x^{1,j} - d)^T P_1 (x - d) \le 1,  \nonumber \\
		& (x^{2,j} - d)^T P_2 (x - d) \le 1,  \nonumber
	\end{align}
\end{subequations}
which are linearly approximated by two quadratic constraints \eqref{4_3}. $x^{1,j}$ and $x^{2,j}$ are the boundary points of 
\begin{subequations}
	\begin{align}
		& (x - d)^TP_1(x - d) \le 1,  \label{4_3_1} \\
		& (x - d)^TP_2(x - d) \le 1,  \label{4_3_2}
	\end{align} \label{4_3}
\end{subequations} respectively,
and
\begin{equation}
	d = \begin{bmatrix}
		-2.15 \\ 0 
	\end{bmatrix}, \ P_1 = \begin{bmatrix}
		0.14 & 0.17 \\ 0.17 & 1.7
	\end{bmatrix}, \
	P_2 = \begin{bmatrix}
		0.20 & 0.05 \\ 0.05 & 0.21	
		\end{bmatrix}. \nonumber
\end{equation}
Similarly, the terminal constraint is constructed by piecewise linear approximation of two quadratic constraints \eqref{4_3_1} and $(x - d)^TP_3(x - d) \le 1$ with $P_3 = \begin{bmatrix}
	0.1 & 0.7 \\ 0.7 & 0.97
\end{bmatrix}$. The state constraints and terminal constraint are shown in Fig. \ref{fig:trajectory}. 

The MPC cost function is defined as \eqref{2_4} where
\begin{equation}
	Q = \begin{bmatrix}
		1 & 0 \\ 0 & 1
	\end{bmatrix}, \ P = \begin{bmatrix}
	1 & 0 \\ 0 & 1
	\end{bmatrix}, R = 1. \nonumber
\end{equation}
The prediction horizon is $N = 12$ and the total number of constraints is $n_c = 1948$. The Lipschitz constant $\hat{\kappa}_{\text{max}} = 21.44$  is calculated by \eqref{3_6} where
\begin{equation}
	\Phi = \text{diag}(\frac{1}{\Vert G_1 \Vert_2}, \frac{1}{\Vert G_2 \Vert_2}, \cdots, \frac{1}{\Vert G_{n_c} \Vert_2}).\nonumber 
\end{equation}

\begin{figure}[t]
	\centering
	\includegraphics[width=8cm]{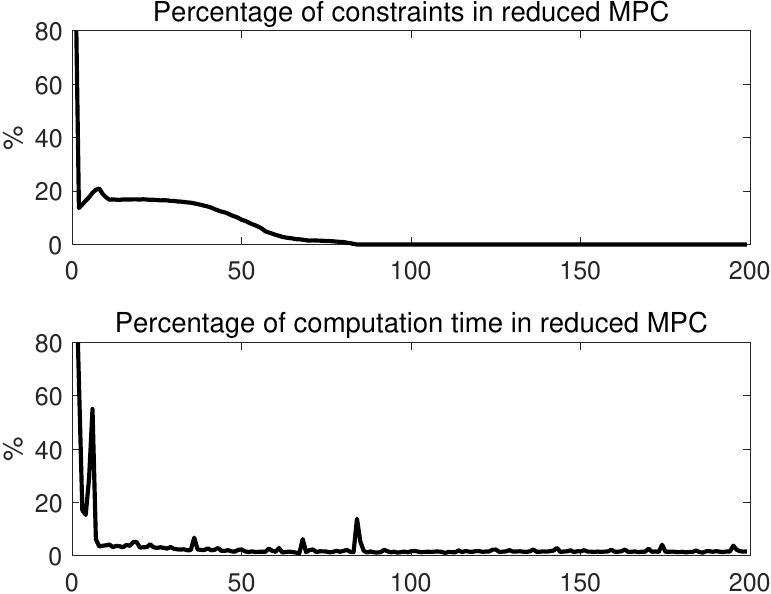}
	\caption{The percentage of constraints and the computation time with respect to the original MPC solution over time.}
	\label{fig:percentage}
\end{figure}

The system is initialized at $x_0 = \begin{bmatrix} -4.1 & 0 \end{bmatrix}^T $ and expected to converge to the origin under the control policy. The data set $D$ is initialized with one data $(0,0,\emptyset)$. The resulting state trajectories are shown in Fig. \ref{fig:trajectory}. The closed-loop trajectories under the original MPC and the proposed ca-MPC steer towards the origin without violating the constraints. Moreover, our proposed ca-MPC produces the same trajectories as that of the original MPC. The comparation of the constraints number and computation time between two schemes are shown in Fig. \ref{fig:percentage}. After 10 steps, the proposed scheme decreases more than 80\% constraints, then the proposed scheme removes almost all of constraints after 80 steps. In addition, our ca-MPC scheme is 10-100 times faster to compute compared to the original MPC scheme.

\section{Conclusions} \label{con}

In this paper, we presented the constraint-adaptive MPC scheme which can exploit the historical data to remove constraints. A crucial aspect of our method is to analyze the Lipschitz continuity and design the outer approximation. In particular, we provided an explicit form of the Lipschitz constant which can be calculated offline and efficiently. Then, we proved that 
the closed-loop behavior remains unchanged compared to the original MPC policy. Simulations validated that our scheme can significantly reduce the number of constraints and improve the computational speed.
%

\begin{appendices} \label{app}
\section{Proof of Theorem \ref{continuity}}
	The KKT conditions related to the mp-QP \eqref{3_2} are
\begin{subequations}
	\begin{align}
		Hz + F^Tx + G_\mathbb{I}^T \lambda = 0, \label{A_1_1} \\
		\lambda \circ (G_\mathbb{I}z - W_\mathbb{I} - S_\mathbb{I} x) = 0, \label{A_1_2} \\
		\lambda \ge 0, \label{A_1_3} \\
		G_\mathbb{I}z \le W_\mathbb{I} + S_\mathbb{I}x. \label{A_1_4}
	\end{align} \label{A_1}
\end{subequations}
Solving \eqref{A_1_1} for z yields
\begin{equation}
	z = -H^{-1}(F^Tx + G_\mathbb{I}^T\lambda). \label{A_2}
\end{equation}
Let superscript $\tilde{}$ denotes the variable corresponding to active constraints $\mathcal{A}_\mathbb{I}(x)$. Without loss of generality, we assume that the rows of $\tilde{G}$ are linearly independent \cite{bemporad2002explicit}. For active constraints, $\tilde{G}_\mathbb{I}z - \tilde{W}_\mathbb{I} - \tilde{S}_\mathbb{I} x = 0$. Combining it with \eqref{A_2}: 
\begin{equation}
	\tilde{\lambda} = - (\tilde{G}_\mathbb{I} H^{-1} \tilde{G}_\mathbb{I}^T)^{-1}(\tilde{W}_\mathbb{I} + (\tilde{S}_\mathbb{I} + \tilde{G}_\mathbb{I}H^{-1}F^T)x) \label{A_3}
\end{equation}
where $\tilde{G}_\mathbb{I}, \tilde{W}_\mathbb{I}, \tilde{S}_\mathbb{I}$ correspond to the set of active constraints, and $\tilde{G}H^{-1}\tilde{G}^T$ is invertible because the rows of $\tilde{G}$ are linearly independent. Substituting \eqref{A_3} into \eqref{A_2}:
\begin{align*}
	z^*_\mathbb{I}(x) = & -H^{-1}F^Tx \\ 
	& + H^{-1}\tilde{G}_\mathbb{I}^T (\tilde{G}_\mathbb{I} H^{-1} \tilde{G}_\mathbb{I}^T)^{-1}(\tilde{W}_\mathbb{I} + (\tilde{S}_\mathbb{I} + \tilde{G}_\mathbb{I}H^{-1}F^T)x). \label{A_4}
\end{align*}
Define
\begin{equation}
	K = -H^{-1}F^T +  H^{-1}\tilde{G}_\mathbb{I}^T (\tilde{G}_\mathbb{I} H^{-1} \tilde{G}_\mathbb{I}^T)^{-1}(\tilde{S}_\mathbb{I} + \tilde{G}_\mathbb{I}H^{-1}F^T). \nonumber
\end{equation}
We obtain
\begin{equation}
	\begin{split}
		\Vert K \Vert_2 \le & \Vert H^{-1} F^T\Vert_2 \\
		& + \Vert H^{-1}\tilde{G}_\mathbb{I}^T (\tilde{G}_\mathbb{I} H^{-1} \tilde{G}_\mathbb{I}^T)^{-1}(\tilde{S}_\mathbb{I} + \tilde{G}_\mathbb{I}H^{-1}F^T) \Vert_2. \label{A_6}
	\end{split}
\end{equation}

In fact, $\tilde{G}_\mathbb{I}, \tilde{S}_\mathbb{I}$ are the submatrices composed by the rows in $G, S$ respectively. We can represent these matrix with $J$ such that
\begin{equation}
	\tilde{G}_\mathbb{I} = JG, \ \tilde{S}_\mathbb{I} = JS. \nonumber
\end{equation} 
Here, the matrix $J$ is non-square matrix that has at most one entry of 1 in each row and each column with all other entries 0. Substituting this into the second term of \eqref{A_6} yields
\begin{equation}
	\begin{aligned}
		&\Vert H^{-1}\tilde{G}_\mathbb{I}^T (\tilde{G}_\mathbb{I} H^{-1}  \tilde{G}_\mathbb{I}^T)^{-1}(\tilde{S}_\mathbb{I} + \tilde{G}_\mathbb{I}H^{-1}F^T) \Vert_2 \\
		& = \Vert H^{-1}G^TJ^T (JG H^{-1} G^TJ^T)^{-1}(JS + JGH^{-1}F^T) \Vert_2 \\
		& \le \Vert H^{-1}G^T \Vert_2 \Vert J^T (JG H^{-1} G^TJ^T)^{-1}J \Vert_2 \Vert S + GH^{-1}F^T\Vert_2. 
	\end{aligned} \label{A_8}
\end{equation}
The last inequation is built with the submultiplicative property of the spectral norm. It follows that
\begin{equation}
	\begin{aligned}
		& \Vert J^T (JG H^{-1} G^TJ^T)^{-1}J \Vert_2\\
		& \le \Vert J^T \Vert_2 \Vert (JG H^{-1} G^TJ^T)^{-1} \Vert_2  \Vert J \Vert_2\\
		& = \Vert (JG H^{-1} G^TJ^T)^{-1} \Vert_2 = \lambda_{1}((JG H^{-1} G^TJ^T)^{-1}) \\
		& = \frac{1}{\lambda_{\text{min}}(JG H^{-1} G^TJ^T)},
	\end{aligned} \label{A_9}
\end{equation}
where the second and third equation are satisfied because $(JG H^{-1} G^TJ^T)^{-1}$ is positive definite matrix. 

$H^{-1}$ is positive definite matrix and can be written as $H^{-1} = LL^T$ (i.e. Cholesky decomposition) where $L$ has full rank. Let $Y=GL$ The matrices $JGLL^TG^TJ^T$ and $L^TG^TJ^TJGL$ have the same non-zero eigenvalues and the number of non-zero eigenvalues is $\mathcal{A}(x)$:
\begin{equation}
	\begin{split}
		&\lambda_{\text{min}}(JG H^{-1} G^TJ^T)  = \lambda_{\text{min}}(JYY^TJ^T) \\ 
		&= \lambda_{\text{card}(\mathcal{A}(x))}(Y^TJ^TJY) = \lambda_{\text{card}(\mathcal{A}(x))}(\sum_{j \in \mathcal{A}(x)} Y^T_j Y_j). \label{A_10}
	\end{split}
\end{equation}

According to the Corollary in \cite[Corollary 4.3.15.]{horn2012matrix}, if two matrices $A, B$ are Hermitian, then $\lambda_i(A) + \lambda_1(B) \le \lambda_i(A+B)$. As a result,
\begin{equation}
	\begin{split}
		&\lambda_{\text{card}(\mathcal{A}(x))}(\sum_{j \in \mathcal{A}(x)} Y^T_j Y_j) \\
		& \ge \lambda_{\text{card}(\mathcal{A}(x))}(\sum_{j \in \mathcal{A}(x), j \neq j_0} Y^T_j Y_j) + \lambda_1(Y^T_{j_0} Y_{j_0}), \nonumber
	\end{split}
\end{equation}
where $j_0$ can be any element in $\mathcal{A}(x)$. The rank of $\sum_{j \in \mathcal{A}(x), j \neq j_0} Y^T_j Y_j$ is less than $\text{card}(\mathcal{A}(x))$. So its $\text{card}(\mathcal{A}(x))$-th largest eigenvalue is zero. Moreover, $\lambda_1(Y^T_{j_0} Y_{j_0}) = Y_{j_0}Y^T_{j_0}$.  We have
\begin{equation}
	\begin{split}
		& \lambda_{\text{card}(\mathcal{A}(x))}(\sum_{j \in \mathcal{A}(x)} Y^T_j Y_j)  \ge Y_{j_0}Y^T_{j_0} \\ 
		& \ \ \ \ \ge \min_j Y_jY_j^T = \min_j G_jH^{-1}G_j^T. \label{A_12}
	\end{split}
\end{equation}

Combing \eqref{A_8},\eqref{A_9},\eqref{A_10},\eqref{A_12} with \eqref{A_6} leads to
\begin{equation}
	\begin{split}
		\Vert K \Vert_2 & \le \Vert H^{-1} F^T\Vert_2 \\
		& + \frac{1}{\min_j G_jH^{-1}G_j^T}\Vert H^{-1}G^T \Vert_2 \Vert S + GH^{-1}F^T\Vert_2, \nonumber 
	\end{split}
\end{equation}
which completes the proof.

\end{appendices}

\bibliographystyle{IEEEtran}
\bibliography{IEEEabrv,reference}

\end{document}